\documentclass[aps,prl,twocolumn,superscriptaddress,showpacs]{revtex4}
\usepackage{graphicx}

\begin{document}
\bibliographystyle{apsrev}
\title{Signature of Magnetic Phase Separation in the Ground State of
$\rm Pr_{1-x}Ca_{x}MnO_3$}

\author{Hao Sha}
\affiliation{Department of Physics, Florida International
University, Miami, Florida 33199}

\author{F. Ye}
\affiliation{Neutron Scattering Science Division,
Oak Ridge National Laboratory, Oak Ridge, Tennessee 37831-6393}

\author{Pengcheng Dai}
\affiliation{Department of Physics and Astronomy,
The University of Tennessee, Knoxville, Tennessee 37996-1200}
\affiliation{Neutron Scattering Science Division,
Oak Ridge National Laboratory, Oak Ridge, Tennessee 37831-6393}

\author{J. A. Fernandez-Baca}
\affiliation{Neutron Scattering Science Division,
Oak Ridge National Laboratory, Oak Ridge, Tennessee 37831-6393}

\affiliation{Department of Physics and Astronomy,
The University of Tennessee, Knoxville, Tennessee 37996-1200}

\author{D.~Mesa}
\affiliation{Department of Physics, Florida International
University, Miami, Florida 33199}

\author{J. W.~Lynn}
\affiliation{NIST Center for Neutron Research,
Gaithersburg, Maryland, 20899-6102}

\author{Y. Tomioka}
\affiliation{Correlated Electron Research Center (CERC), National Institute of Advanced Industrial Science and Technology (AIST),
Tsukuba 305-0046, Japan}

\author{Y. Tokura}
\affiliation{Correlated Electron Research Center (CERC), National Institute of Advanced Industrial Science and Technology (AIST),
Tsukuba 305-0046, Japan}
\affiliation{Department of Applied Physics, University of Tokyo,
Tokyo 113-8656, Japan}

\author{Jiandi Zhang}
\email{zhangj@fiu.edu}
\affiliation{Department of Physics, Florida International
University, Miami, Florida 33199}

\date{\today}

\begin{abstract}
Neutron scattering has been used to investigate the evolution of the
long- and short-range charge-ordered (CO), ferromagnetic
(FM), and antiferromagnetic (AF) correlations in single crystals
of $\rm Pr_{1-x}Ca_{x}MnO_3$. The existence and population of spin clusters
as reflected by short-range correlations are found to drastically
depend on the doping ($x$) and temperature (\emph{T}). Concentrated spin clusters coexist with
long-range canted AF order in a wide temperature range in $x$ = 0.3
while clusters do not appear in $x$ = 0.4 crystal. In contrast, both CO and
AF order parameters in the $x$ = 0.35 crystal show a precipitous decrease
below $\sim$ 35 K where spin clusters form. These results provide
direct evidence of magnetic phase separation and indicate that there
is a critical doping $x_{c}$ (close to $x$ = 0.35) that divides the
phase-separated site-centered from the homogeneous bond-centered
or charge-disproportionated CO ground state.
\end{abstract}


\maketitle
The $R_{1-x}A_xMnO_3$ (where $R$ and $A$ are rare- and
alkaline-earth ions) manganites, known for the existence of ``colossal''
magnetoresitance, provide an archetype laboratory for the study of
phase inhomogeneities \cite{dagottobook}. In general, doping ($x$)
not only causes a quenched disorder due to the size mismatch of
$A$-site cations in the perovskite lattice, but also
induces a distinctive chemical valence in Mn ions (Mn$^{3+}$
vs.~Mn$^{4+}$ in the original portrayal \cite{wollan55,goodenough55}).
These two kinds of Mn ions can assemble themselves into a CE-type checkerboard-like
charge ordered (CO) state at low temperature (LT) as shown in Fig. 1(b), promoting an antiferromagnetic (AF) insulating phase that
competes with the ferromagnetic (FM) metallic state \cite{jirak80,tokura96,tomioka96}.
In a conventional CO state, referred to as a site-centered CO structure, distinct
Mn$^{3+}$ and Mn$^{4+}$
ions in half-doped manganites ($x$ = 0.5) form 1:1 ratio. However, another possible CO pattern may also occur, namely
the bond-centered structure in which the charge is localized
not on Mn sites but on Mn-O-Mn bonds with no distinctive Mn$^{3+}$/
Mn$^{4+}$ sites \cite{aladine02,ferrari03,zhen02,zener51}, thus raising an unsettled
issue about the nature of the ground state around $x$ = 0.5.

Furthermore, even in some non half-doped ($x$ $\neq$ 0.5) manganites,
an ostensible {\it pseudo-}CE-type CO state has been reported
\cite{tomioka96,pollert82,tokura06,cox98}. Within the CE-type CO
frame, excess of electronic charge (or Mn$^{3+}$ ions) with respect
to the ideal half-doped ($x$ = 0.5) case exists in these non
half-doped compounds. Consequently, important issues naturally
emerge: how is the excessive electronic charge distributed in the CO
state and what are the arrangements of spins and orbitals?
There are at least three possible scenarios based on the charge
distribution. First, the excessive electronic charge is
distributed locally in the CE-type motif
with distinct Mn$^{3+}$ and Mn$^{4+}$ sites. Apparently an
electronic/magnetic phase separation is unavoidable due to unequal
amount of Mn$^{3+}$ and Mn$^{4+}$ ions while keeping the rigid
CE-type Mn$^{3+}$ and Mn$^{4+}$ order. Another one is that the
excessive charge is distributed uniformly in the CE-type motif with certain charge
disproportionation($\delta$) \cite{brink99,grenier04}. Both scenarios can be categorized
as the site-centered structure because of having distinctive Mn
sites, one inhomogeneous while the other homogeneous. The third
scenario is the completely homogenous bond-centered structure
in which all Mn sites are equivalent ($\delta$
= 0) \cite{aladine02,jooss07}. Presently it is unclear which scenario
is more appropriate for the observed CO state. Alternatively, could
it be possible that the CO structure depends on doping concentration?

In this letter, we report the signature of magnetic phase separation
in the CO state from neutron scattering studies of a prototype
manganite system: $\rm Pr_{1-x}Ca_{x}MnO_3$ (PCMO).  Our experimental
results suggest a critical doping concentration $x_{c}$ that divides
homogeneous and inhomogeneous CO ground state.  As shown in Fig.~1
(a), the non-metallic CO ground state of PCMO
exists over a broad doping range ($0.3 \leq x \leq 0.7$)
\cite{tomioka96}. Because Pr$^{3+}$ and
Ca$^{2+}$ have almost equal ion radii, PCMO has negligible
quenched disorder, and thus is an ideal system to elucidate how doping
affects the structure of CO and magnetic ground states.

\begin{figure}[ht!]
\includegraphics[width=2.9in]{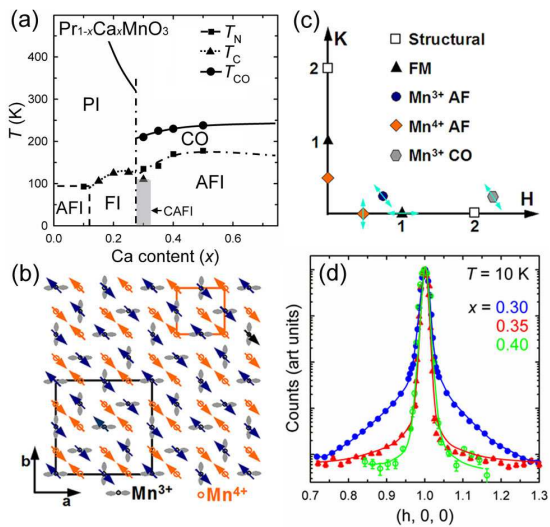}
\caption{\label{fig:fig1}
(a) Generic phase diagram of $\rm Pr_{1-x}Ca_xMnO_3$
\cite{tomioka96}.  (b) Schematic $ab$-plane view of the CE-type
structure with both orbital $(d_{3x^2-r^2}/d_{3y^2-r^2)}$ and spin
ordering. The bigger square denotes the Mn$^{3+}$ orbital/magnetic
unit cell while the smaller square shows the Mn$^{4+}$ magnetic unit
cell. (c) The probed superlattice peak positions and the scan
directions (marked by arrows) in reciprocal space. (d)
Normalized $q$-scan profiles of the scattering near (1, 0, 0) at
$T$ = 10 K for PCMO30 (blue), PCMO35 (red), and PCMO40 (green). The
solid lines are guides to the eye.}
\end{figure}

Three single crystals with $x$ = 0.3 (PCMO30), 0.35 (PCMO35),
and 0.4 (PCMO40) were grown by the floating-zone method
\cite{tomioka96}.  The measurements were carried
out using the HB-1 and HB-3 triple-axis spectrometers at the High
Flux Isotope Reactor at the Oak Ridge National Laboratory (ORNL),
BT-7, and BT-9 at the NIST Center for Neutron Research. For
simplicity, we label all wave vectors in terms of the pseudo-cubic
unit cell with lattice parameters of $a$ = 3.87 $\AA$, although all
of our samples have orthorhombic structures slightly distorted from
the cubic lattice. The wave vector $\emph{\textbf{Q}} = (Q_x, Q_y, Q_z)$ is
in the unit of $\AA^{-1}$ and $(H, K, L) = (Q_xa/2\pi, Q_ya/2\pi,
Q_za/2\pi)$ is in reciprocal lattice units (rlu). The samples
were aligned to allow the wave vector in the form of $(H, K, 0)$
accessible in the horizontal scattering plane.

All of the measurements were based upon the
CE-type structure with CO and AF ordering shown in Fig.~1(b).
Including the orbital part, the periodicity of AF order for
Mn$^{3+}$ spins is twice of that for Mn$^{4+}$ spins. As shown in
Fig.~1(c)\cite{ye05}, we used $\emph{\textbf{Q}}$ = (2.25, 0.25, 0),
(1, 0, 0), (0.75, 0.25, 0), and (0.5, 0, 0), to probe the order of
the CO, FM, AF at Mn$^{3+}$ and Mn$^{4+}$ sites, respectively.
The corresponding short-range correlations or, alternatively, clusters,
were also determined by measuring the diffuse scattering around these characteristic
positions \cite{analysis}. As listed in Table 1, the measured transition temperatures
for CO, AF, and FM phase are in agreement with those reported in the
literature. A canted long-range FM order coexists with an AF structure
in PCMO30 below $T_C$ = 110~K (see the inset (a) of Fig.~2), known as a
canted AF insulating (CAFI) phase \cite{yoshizawa95}.  However, no
long-range FM order was detected in the ground state for either
PCMO35 or PCMO40.

\begin{table} [ht!]
\begin{tabular}{lccc}
\hline
\hline
$\rm Pr_{1-x}Ca_{x}MnO_3$ & $x=0.30$ & $x=0.35$ & $x=0.40$ \\
\hline
$T_{CO}; T_N; T_C$ (K) & 210;125;110 & 230;160;N/A & 245;170;N/A \\
$\rho_{FM}(\times 100\%)$ & $15.1\pm2.0$ & N/A &  N/A \\
$\rho_{AF}(\times 100\%)$ & $28.7\pm0.8$ & $0.160\pm0.003$ &  N/A \\
\hline
\hline
\end{tabular}
\caption{\label{tab:tab1}
The integrated intensity ratios of the diffuse and total
(diffuse plus Bragg) magnetic scattering
($\rho_{FM,AF}\equiv {I_{diffuse}(FM,AF)}/{I_{total}(FM,AF)}$) vs. $x$ measured at 10 K:
$\rho_{FM}$, the ratio obtained from the scan across (1, 0,
0) by subtracting the lattice contribution to the Bragg peak;
$\rho_{AF}$, the ratio obtained from the scan across (0.5, 0, 0).}
\end{table}

We observed a strong doping dependence of both AF and FM diffuse scattering at
the ground state. Figure 1(d) displays the normalized \emph{q}-scans at (1, 0, 0)
including both magnetic and structural scattering from the three
doping levels.  Similar to that reported before
\cite{yoshizawa95,kajimoto98}, a strong diffuse component appears
near the Bragg peak for PCMO30, which indicates that FM spin
clusters coexist with the long-range FM order in the ground state.
In contrast, PCMO35 has a rather weak FM diffuse shoulder. PCMO40
shows only a nice Gaussian profile due to long-range lattice
scattering with no sign of short-range spin correlation. The AF
scattering profiles at (0.5, 0, 0) and (0.75, 0.25, 0) reveal
a similar $x$-dependence of diffuse components. As summarized in
Table 1, PCMO30 has a CAFI ground state with a significant amount of
FM/AF clusters ($\rho_{FM} \approx 15.1 \pm 2.0 \%$ and $
\rho_{AF} \approx 28.7 \pm 0.8 \%$). PCMO35 has an almost homogeneous AF ground
state with a very small weight of spin clusters, and PCMO40 has a
completely uniform AF ground state.  Therefore, a critical doping
$x_{c}$ must exist and be very close to the value of
$x$ = 0.35, which serves as the boundary between the homogeneous and
inhomogeneous ground state of $\rm Pr_{1-x}Ca_{x}MnO_3$.

\begin{figure}[ht!]
\includegraphics[width=2.9in]{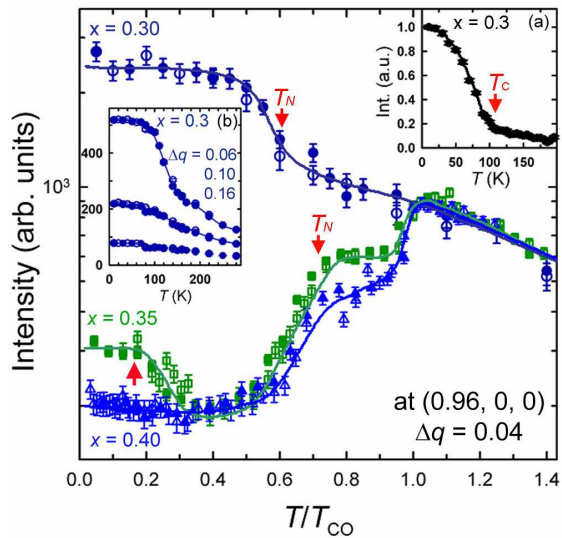}
\caption{\label{fig:fig2}
$T$-dependence of the FM diffuse component measured at (0.96, 0, 0)
($\bigtriangleup$$q$ = 0.04 rlu) for different $x$'s of $\rm
Pr_{1-x}Ca_xMnO_3$ on cooling (solid symbols) and warming (open
symbols). The solid curves are guides to the eye. Arrows mark the
Curie and N\'{e}el temperatures, as well as the onset of spin phase
separation for PCMO35. The insets present the $\it T$-dependence of (a)
the FM order parameter (Bragg peak intensity)
and (b) the FM diffuse component at selected  \emph{q} positions
from (1, 0, 0) for PCMO30.}
\end{figure}

To further reveal such a critical doping behavior for the
inhomogeneity in the magnetic structure, we have examined the
$\it T$-dependence of the FM diffuse scattering intensity at
$\emph{\textbf{Q}}$ = (0.96, 0, 0), which is outside the influence
of the Bragg peak for long-range ordering.  Figure 2 presents
the measured intensity as a function of $\it T$/$T_{CO}$
for all three doping levels.  Similar behavior is obtained for
different positions ( see the inset (b) of Fig.~2).  The FM diffuse component
clearly displays a different $T$-dependence above and below $T_{CO}$.
Comparable results have also been reported by Kajimoto {\it et al.}
\cite{kajimoto98}. When $T > T_{CO}$, all three doping levels of
crystals exhibit a similar $T$-dependence of the diffuse component,
reflecting the FM spin fluctuations in the paramagnetic phase, which are
presumably induced by double-exchange (DE) interaction \cite{dagottobook}.

On the other hand, a completely different $T$-dependence of the diffuse
components for different doping levels appears below $T_{CO}$. For
PCMO40, the FM spin fluctuations deteriorate below $T_{CO}$ and
vanish below $T_N$.  In principle, such a $T$-dependence in PCMO40
can be understood as follows. When $T < T_{CO}$, the DE-mediated
FM spin fluctuations will be suppressed by the charge localization. When $T < T_N$, the FM
fluctuations diminish further because of the onset of AF
order.  For PCMO35, the FM spin fluctuations exhibit the same
$T$-dependence as those in PCMO40, thus indicating that they
originate from the same mechanism.  However, spin clusters
appear below $\sim$ 65 K with both FM (see Fig.~2) and AF characters, as
reflected by the corresponding diffuse components.
This is distinctive from the FM spin fluctuations
at high temperature. In sharp contrast with those in both PCMO35 and
PCMO40, the population of spin clusters in PCMO30, reflected by the
intensity of FM diffuse component, increases rather than decreases
below $T_{CO}$. Surprisingly, it increases much more drastically
below $T_N$, regardless of the establishment of AF order, thus
suggesting the spin clusters in the ground state may have a
different nature from the DE-mediated FM fluctuations at high
temperature.

\begin{figure}[ht!]
\includegraphics[width=2.9in]{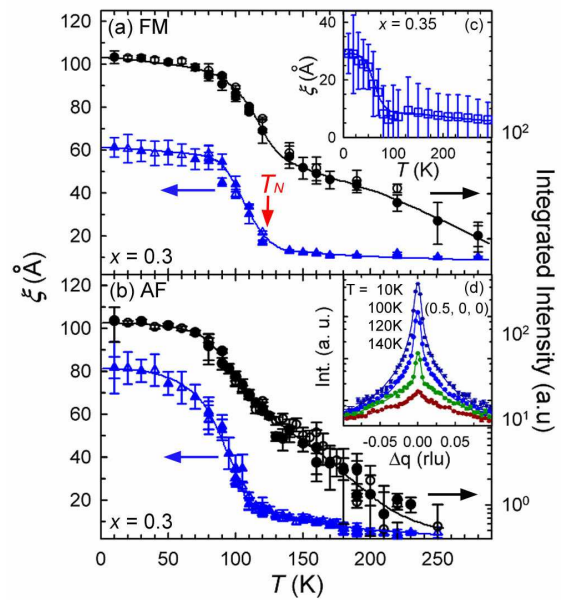}
\caption{\label{fig:fig3}
$\it T$-dependence of the intensity and extracted short-range
correlation length of (a) FM at (1, 0, 0) and (b) Mn$^{4+}$ AF
at (0.5, 0, 0) in PCMO30. The solid lines are
guides to the eye. The solid symbols are for cooling while open
symbols for warming. The insets show the $\it T$-dependence of (c) FM
short-range correlation length in PCMO35 with large error bars due
to the weak diffuse scattering profiles and (d){\it q}-scans of Mn$^{4+}$ AF at
(0.5, 0, 0).}
\end{figure}

The spin clusters appearing in PCMO30 in the LT regime also display
both FM and AF characters, similar to those existing in the ground
state of PCMO35. The integrated diffuse
scattering intensity and the extracted short-range correlation
length $\xi$ near the FM and AF peak exhibit very similar $T$-dependence (see Fig.~3).
The measured average cluster diameter ({\it i.e.} correlation length) is $\xi_{FM}
\approx 60$ $\rm \AA$ from FM and $\xi_{AF} \approx 80$ $\rm \AA$ from AF
scattering. Using a simple estimated population relation in the {\it ab}-plane
of the crystal: $\rho_{FM,AF} \propto n_{FM,AF}\xi^2_{FM,AF}$,
where $n_{FM,AF}$ is the in-plane cluster density,
we find that $n_{FM} \cong n_{AF}$ with measured $\rho_{FM,AF}$ (Table 1), regardless of
the difference in correlation length. Therefore, we speculate that the
measured AF and FM diffuse scattering may be indeed caused from the
same assembly of spin clusters.

Due to the fairly weak diffuse scattering and uncertain correlation
length (see the inset (c) of Fig.~3) we were not able to do a similar
estimation for PCMO35.  Nevertheless, the ground state in both
PCMO30 and PCMO35 is a phase-separated state containing spin
clusters embedded in either an AF or canted AF ordered matrix. The main
difference between these two doping levels is that the magnetic
phase separation exists in the entire measured temperature range in PCMO30
but appears with much smaller population only at LT in PCMO35. One
can anticipate that such a phase-separated ground state will
disappear in a crystal of $\rm Pr_{1-x}Ca_xMnO_3$ with a doping
level slightly larger than $x$ = 0.35. Such an evolution of phase
separation with doping may provide insight into the doping
dependence of the observed CMR effect \cite{tomioka96}.  This
evolution may also explain why a smaller critical magnetic field is
sufficient to melt the CO state for PCMO30 than that for PCMO35 and
PCMO40.

\begin{figure}[ht!]
\includegraphics[width=2.9 in]{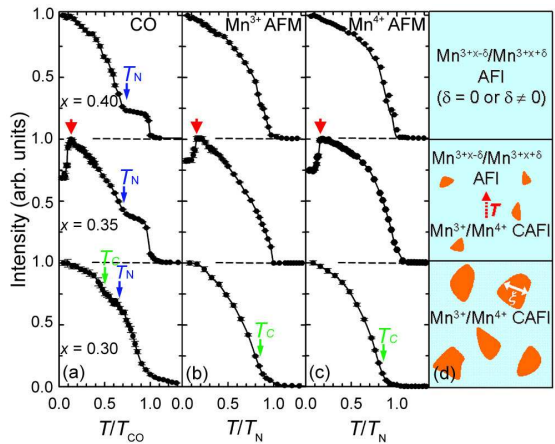}
\caption{\label{fig:fig3}
$T$-dependence of order parameters for (a) CO structure measured at
(2.25, 0.25, 0), (b) Mn$^{3+}$ AF ordering at
(0.75, 0.25, 0), and (c) Mn$^{4+}$ AF ordering at (0.5, 0, 0), for
the three doping $\it x$'s of $\rm Pr_{1-x}Ca_xMnO_3$.
Arrows mark the Curie and N$\acute{e}$el temperatures as well as the onset
of phase separation for PCMO35. The column (d) illustrates the evolution
from an inhomogeneous site-centered to a homogeneous CO phase for
corresponding doping levels.}
\end{figure}

If indeed PCMO35 is a system which undergoes a phase evolution from
homogeneous to inhomogeneous spin-ordered state on cooling, then the
appearance of spin clusters at LT should affect the long-range order
parameters. To elucidate this issue we have investigated
systematically the order parameters of the CO and AF at both Mn$^{3+}$
and Mn$^{4+}$ sites, respectively. As clearly shown in Fig.~4, an anomaly
characterized by a sudden drop in the intensity of the measured order
parameters in PCMO35 emerges at $\sim$ 35 K, coinciding with the establishment of
spin clusters and thus, providing a clear signature of
magnetic phase separation. In addition, we note that the
line shape of these order parameters in PCMO30 near both $T_{CO}$
and $T_N$ is slightly different from that in PCMO35 and PCMO40. The
transitions in PCMO35 and PCMO40 show more pronounced critical
behavior than those in PCMO30. The nature behind this should be
associated with the existence of phase separation near $T_N$ and
$T_{CO}$ in PCMO30, which does not occur in the other two doping levels.

In summary, we have observed a strong $x$- and $T$-dependence of
magnetic phase separation in $\rm Pr_{1-x}Ca_xMnO_3$ crystals.
Spin clusters with both AF and FM correlations coexist with a
CAFI structure in PCMO30, suggesting an inhomogeneous CO state below
$T_{CO}$. The ground state of PCMO30 is inhomogeneous site-centered
($\delta >0 $) or even in the extreme case, the inhomogeneous
CE-type [see Fig. ~4(d)].  In contrast, the observed uniform AF
ordered structure suggests a homogeneous AF CO state in PCMO40.
We have identified a critical doping $x_{c}$, which
is very close to $x$ = 0.35, to divide homogeneous and
inhomogeneous CO ground states.  Whether the homogenous CO state
exists in PCMO40 as a site-centered type with charge
disproportionation ($\delta \neq 0$) or a bond-centered
type is still unclear. One possible
experimental method to distinguish these two types of CO structures
is resonant x-ray scattering \cite{abbamonte06}. Yet, it is clear
that in a manganite away from half doping, a CE-type Mn$^{3+}$/Mn$^{4+}$ CO
state can not survive without phase separation.

This work was supported by U.S. DOE FG02-o4ER46125 and NSF
DMR-0346826.  P.D. was supported by U.S. NSF DMR-0756568. ORNL is
managed by UT-Battelle, LLC, for the U.S. DOE DE-AC05-00OR22725.


\begin{references}
\bibitem{analysis}The diffuse components of scattering were fitted to Lorentzian lineshapes convoluted with
the instrumental resolution. The linewidth $\Gamma$ corresponds to the  inverse correlation length ($\Gamma  =  1 / \xi$), where $\xi$ is also the spin cluster size.  The intensity of the Lorentzians is proportional to the population of spin clusters.
\end{references}
\end{document}